\documentclass[iop]{emulateapj}
\usepackage{natbib}
\usepackage{graphicx}
\usepackage{footnote}
\usepackage{amssymb}

\received{}
\revised{}
\accepted{}

\shorttitle
{
TDG in the Leo Triplet?
}
\shortauthors{B. Nikiel-Wroczy\'nski et al.}

\begin{document}

\title{Discovery of a tidal dwarf galaxy in the Leo Triplet}
\author{B. Nikiel-Wroczy\'nski}
\affil{Obserwatorium Astronomiczne Uniwersytetu Jagiello\'nskiego,
ul. Orla 171, 30-244 Krak\'ow, Poland}
\email{iwan@oa.uj.edu.pl}
\author{M. Soida}
\affil{Obserwatorium Astronomiczne Uniwersytetu Jagiello\'nskiego,
ul. Orla 171, 30-244 Krak\'ow, Poland}
\email{soida@oa.uj.edu.pl}
\author{D.J. Bomans}
\affil{Astronomisches Institut, Ruhr-Universit\"at Bochum,
Universit\"atsstrasse 150, 44801 Bochum, Germany}
\email{bomans@astro.rub.de}
\author{M. Urbanik}
\affil{Obserwatorium Astronomiczne Uniwersytetu Jagiello\'nskiego,
ul. Orla 171, 30-244 Krak\'ow, Poland}
\email{urb@oa.uj.edu.pl}

\begin{abstract}

We report discovery of a dwarf galaxy in the Leo Triplet.
Analysis of the neutral hydrogen distribution shows that it rotates
independently of the tidal tail of NGC 3628, 
with a radial velocity gradient of 35--40\,km s$^{-1}$ over
approximately 13\,kpc. The galaxy has a very high neutral
gas content, explaining large part of its total dynamic mass
-- suggesting a small dark matter content. As it is located
at the tip of the gaseous tail, this strongly suggests its tidal
origin. Should it be the case, it would be one of the most 
 confident and closest (to the Milky Way) detections of a tidal dwarf galaxy and,
at the same time, a most detached from its parent galaxy ($\approx$140 \,kpc)
object of this type.

\end{abstract}

\keywords
{
Galaxies: groups: Arp 317, Leo Triplet --
Galaxies: interactions --
Intergalactic medium 
}

\section{Introduction}

The idea of dwarf objects forming from the tidal debris left by
galaxy mergers was first proposed by Zwicky~\cite{zwicky}, who
suggested that interactions in systems of multiple galaxies can
lead to an ejection of the tidal material and formation of an
intergalactic structure, possibly even a dwarf galaxy. However,
the "recycled" galaxies did not achieve much attention, apart
from a symposium talk by Schweizer~\cite{schweizer}. The first object
of this type was discovered by Mirabel et al.~\cite{mirabel}, who
presented a photometric study of the Antennae galaxies, showing
a tidal dwarf galaxy (TDG) formed from the collisional debris.
Since then, many similar objects have been detected -- see eg.
Brinks et al.~\cite{brinks}, or Duc et al.~\cite{vcc2062}.
Recently, Kaviraj et al.~\cite{kaviraj} presented a study of a
sample of 405 nearby TDG candidates, conducting a statistical
analysis of their properties.  Tidal dwarf galaxy candidates have
been found in the Local Volume \citep[within 11 Mpc -- Hunter et
al.][]{Hunter2000}, too. 
The M81 group hosts some of the closest examples of the TDGs. 
Small distance allowed to use the HST-based 
colour-magnitude diagrams \citep[Makarova et al.][]{Makarova2002}
to analyse the star formation history of the TDG candidates 
and search for additional signs of the tidal origin.

What makes the TDGs especially interesting is their mass composition. 
Whereas "normal" galaxies consist mostly of the dark matter (DM), TDGs 
do not; velocity of the DM particles in the galactic halo is much 
higher than the escape velocity of a TDG \citep[Bournaud][]{bournmass},
so they are not kinematically bound to it. Hence, such systems consist
usually of the baryonic matter only. Additionally, as they are formed in 
the outer parts of the galactic disks, their metallicity is higher 
than in the non-tidal dwarfs.

Only several TDGs were estimated to be heavy enough to contain
significant non-baryonic fraction, but usual
estimates suggest DM content at most similar to the baryonic mass
-- far below the typical order of magnitude of difference in
non-tidal dwarf systems \cite[Bournaud][]{bournmass}. Lack of
the DM content and specific environment cause the evolution of the TDGs
to be different from that of typical field galaxies, still to be
studied and described. With the low dark matter content TDGs
should also be more susceptible to the formation of galactic
outflows driven by strong star formation. Alternately,
different mass distribution may lead to a lower overall star
formation and therefore to a low surface brightness nature of
evolved TDG.
                                 
TDGs are interesting not only because of their mass composition,
but also because of their influence on the intergalactic  environment.
Tidal debris can interact with other group members, like in the case
of the Leo Triplet galaxy NGC\,3627,  known for its unusual magnetic
field morphology \citep[Soida et al.][]{3627}. Recently,
We\.zgowiec et al.~\cite{3627dwarf} suggested these peculiarities could be
a result of a past collision with a dwarf galaxy. Thus,  TDGs might
play an important role in the further evolution of their
progenitors.

Galaxy systems with massive tidal tails and/or rings constitute
favourable objects to search for the TDG candidates. One of the
best examples of such objects is the Leo Triplet, a nearby group of
galaxies, known for a large tidal plume extending eastwards from
NGC\,3628.  Originally described by Zwicky~\cite{zwicky}, 
the plume was later confirmed by photographic 
observations by Kormendy\&Bahcall~\cite{korm}. Neutral hydrogen
studies by Rots~\cite{rots} and Haynes et al.~\cite{arecibo}
revealed a thick {\rm H}{\sc i} structure, longer and wider than
its optical counterpart. A detailed analysis of the {\rm H}{\sc
i} distribution \citep[Stierwalt et al.][]{hinew} suggested
numerous candidates for the non-tidal dwarf satellites.

Recently, Nikiel--Wroczy\'nski et al.~\cite{leoeff} presented a
study of the magnetic field in the Triplet. The authors suggested
that the {\rm H}{\sc i} clump at the tip of the tidal tail could
be a TDG. However, as pointed out in most of the TDG studies 
\citep[see eg. Kaviraj et  al.][]{kaviraj} determination if a candidate
is self-gravitating (galaxy), or rather a larger part of the
tidal debris (that will never become a self-bound, independent
object) is crucial.

In this paper we use the archive neutral hydrogen and
optical data to show that the velocity field of the TDG candidate
detected in the Leo Triplet exhibits a velocity gradient
and has a faint optical counterpart. 
These findings strongly
support the idea of its independent rotation, thus confirming its
identification as a galaxy.

\section{Observations and  data reduction}

\subsection{Neutral hydrogen observations}

The 1.41\,GHz spectral data, made with the Very Large Array (VLA)
of the National Radio Astronomy Observatory (NRAO)\footnote{NRAO
is a facility of National Science Foundation operated under
cooperative agreement by Associated Universities, Inc.}  in the
D-array configuration, were taken from the NRAO Data Archive
(Project AB1074, PI: A.Bolatto). Two IFs were set, first at
1.41527\,GHz, second at 1.41761\,GHz, each with a bandwidth of
3.1\,MHz. The corresponding velocity range is
269\,--\,1511\,km\,s$^{-1}$. The total bandwidth of the observations
is 5.5\,MHz and the central frequency of the final data is
1.41645\,GHz. We used only these fields that
contained the TDG candidate. Datasets for each of the pointings used were
reduced following the standard spectral line data calibration
procedure in the Astronomical Image Processing System
(\textsc{aips}). The maps were Briggs weighted to detect faint,
extended emission. The final angular resolution  is 80$\arcsec
\times 80 \arcsec$. Assuming the distance to NGC\,3628 of 12.15
Mpc  (median distance calculated from the Tully-Fisher-determined
values from the NASA Extragalactic Database), this yields the linear
resolution of our maps of some 4.7\,kpc. The velocity resolution
is 20.7\,km\,s$^{-1}$ and the r.m.s. noise level ($\sigma$) of
the final maps is approximately 0.63\,mJy. This corresponds to
the brightness temperature of 0.10\,K and {\rm H}{\sc i} column
density threshold of  $6.9\times 10^{19} \mathrm{cm}^{-2}$ (per
spectral channel).

\subsection{Optical observations}

A search for an optical counterpart requires sensitive
optical and/or near-infrared data. We  checked the Sloan Digital
Sky Survey (SDSS) DR7 data \citep[Abazajian et al.][]{DR7} in  all
filters ({\it u', g', r', i'}, and {\it z'}\/) individually. A very low
surface brightness structure  is present in the {\it g'}\/ image.
To improve on this tentative detection we applied our
stacking/filtering procedure, which increases the detectability of very
low  surface brightness structures \citep[see Miskolczi et
al.][for details]{miskolczi}.  

The resulting processed stack of the images
in the three most sensitive SDSS filter bands {\it g', r'}, and 
{\it i'}\/ is
presented in Fig.~\ref{optical}.

\section{Results}

\subsection{Optical emission distribution}
\label{optic}

\begin{figure}[]
 \resizebox{\hsize}{!}{\includegraphics{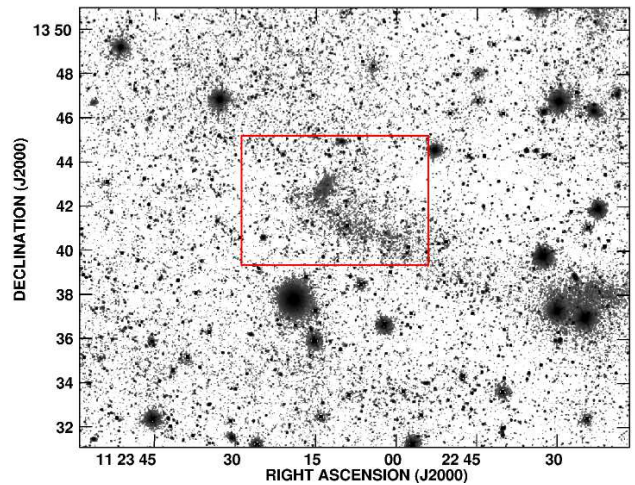}}
\caption{
The processed stack of the SDSS {\it g', r'}, and {\it i'} bands. The faint concentration contained in the central frame is the Leo-TDG. 
}
\label{optical}
\end{figure}

The processed SDSS image stack (Fig.~\ref{optical}) shows a faint
extended region at  the position of the {\rm H}{\sc i}
plume (diffuse, low surface brightness patch at
$\mathrm{R.A._{2000}} =  11^\mathrm{h} 23^\mathrm{m}
15^\mathrm{s}$, $\mathrm{Dec_{2000}} = 13^{\circ}  43\arcmin 15  
\arcsec$) and a (fainter) structure elongated along the E--W
direction.  This structure is also visible in a widefield image
showing the tail of NGC\,3628, provided to us by S.~Mandel
\citep[reproduced in Fig.~4 in Miskolczi et al.][]{miskolczi}. 

The detected patch shows an exponential brightness profile with a central surface brightness of 25.2 
mag sqarcsec$^{-1}$ ({\it g'}\/ filter) and a scale length of $120\arcsec \times 60\arcsec$ 
($7\times 3.5$\,kpc).
The total brightness of the structure is 17.1$^{m}$ (16.65$^{m}$ in the {\it r'}\/ filter). This means that
{\it g'$-$r'} is 0.45$^{m}$. Using the conversion factors by Jester et al.~\citep{jester}, these translate to an 
apparent B-band brightness $m_{\mathrm{B}}$ = 17.45$^{m}$, B$-$V of 0.615$^{m}$, and central surface brightness
$\mu_{\mathrm{B}}$ of 25.55$^{m}$ ($\mu_{\mathrm{V}}$ = 24.93$^{m}$). The distance modulus is 30.42, yielding the absolute B-band magnitude
M$_{\mathrm{B}}$ of $-$12.97$^{m}$.

We also check the colour of the tidal tail at two separate positions, one
closer ($\approx 2\arcmin$ to the SW from the TDG),  one more
distant ($\approx 13\arcmin$ to the SW from the TDG).
The surface brightness of the more distant position in the tidal arm
is comparable to that of the TDG, while the position closer to the
TDG is fainter, which unfortunately limits the accuracy of the
measurement.
Measuring the TDG and both regions in the stream with different
methods of background determination implied that the uncertainty
of the colour measurements of each of these structures is at least 0.1 mag.
With that in mind, the colours of both the distant stream clump ({\it g'}--{\it r'}=0.3) and
the fainter, closer one ({\it g'}--{\it r'}=0.4) are the same as the colour of
the TDG itself. With the data at hand, any difference of colour between the stellar population mix
in the TDG and the two analysed regions in the stream remains within the uncertainties.
A detailed analysis of the stellar populations of the TDG and of the tidal stream must await
much better data. 

\subsection{Neutral hydrogen distribution}
\label{momnt}

\begin{figure}[]
\resizebox{\hsize}{!}{\includegraphics{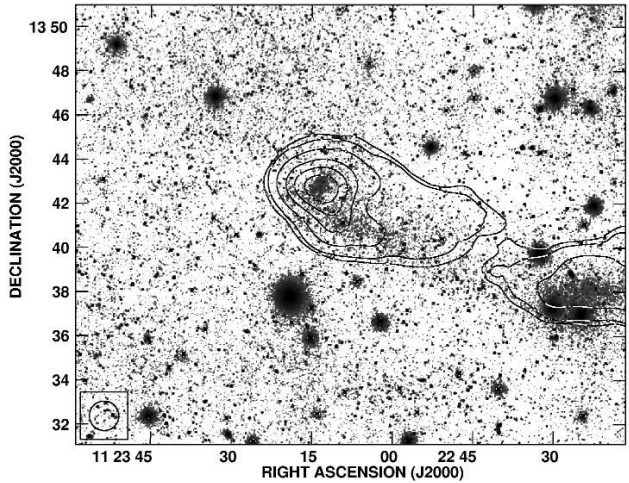}}
\caption{
Contours of the zeroth moment of the {\rm H}{\sc i} distribution from the VLA overlaid on a multi-band ({\it g', r'}, 
and {\it i'}\/) image stack from the SDSS. The contour levels are $5,10,25,50,75 \times 30$ Jy/beam $\times$ m/s.
The angular resolution of the radio data is 80 arcsec.
}
\label{mom0}
\end{figure}

Figure~\ref{mom0} presents the {\rm H}{\sc i} total intensity
(zeroth moment) map of the TDG candidate. It shows a luminous,
well-resolved source of approximately ellipsoidal shape of a
major axis of  300$\arcsec$ and a minor axis of 275$\arcsec$ 
with a position angle of $35\degr$. This corresponds to a linear size of
17.5$\times$16\,kpc. 
The total intensity is $\approx$ 
9.0$\pm$0.5\,Jy\,km\,s$^{-1}$. The neutral hydrogen data have
clear counterparts in the optical regime. Both optical and {\rm
H}{\sc i} emitting media are connected to the tidal tail.

\begin{figure*}[]
\resizebox{\hsize}{!}{\includegraphics{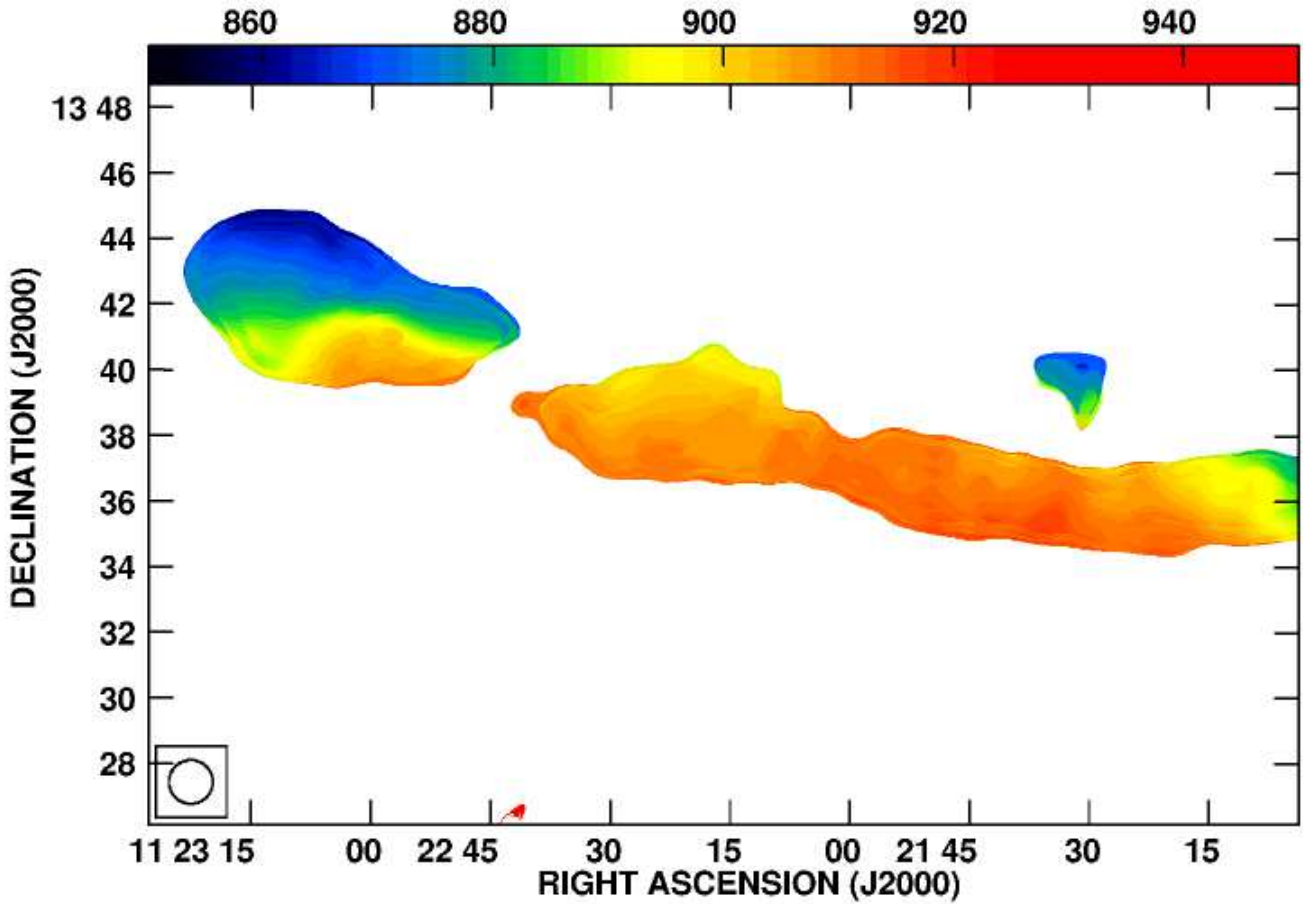}}
\caption{
The {\rm H}{\sc i} velocity distribution map made from the VLA observations. The colours
represent gas species of velocities ranging from 850 (dark navy) to 950\,km\,s$^{-1}$ (red).
The angular resolution is 80 arcsec, and the spectral resolution is 20.7\,km\,s$^{-1}$.
}
\label{mom1}
\end{figure*}

The first moment (velocity) map is shown in Fig.~\ref{mom1}. The
velocity gradient runs from the northern (approaching) to the
southern (receding) side, where it sinks into the tidal plume.
The measured values of the radial velocity range from 
$860 - 900$\,km\,s$^{-1}$. The tail's velocity is somewhat
higher, with a  mean of 910$\pm$ 5\,km\,s$^{-1}$. There is no
observable trend along the E--W direction. As the
velocities of the  dwarf system and the tail are different, it
appears likely that the dwarf is not bound to the tail. This
identifies the dwarf as a separate object, which is
self-gravitating -- \emph{thus, a galaxy.}
This claim is also supported by the morphology of the tidal tail, 
which bends strongly in the direction of the TDG candidate in its
close vicinity. Such a behaviour suggests that the TDG's 
gravitational influence on the tail is higher than expected for 
the internal gravity of the tail. This makes the existence of a 
self-gravitating object in the tip of the tidal tail even more likely.

\begin{figure}[]
\resizebox{\hsize}{!}{\includegraphics{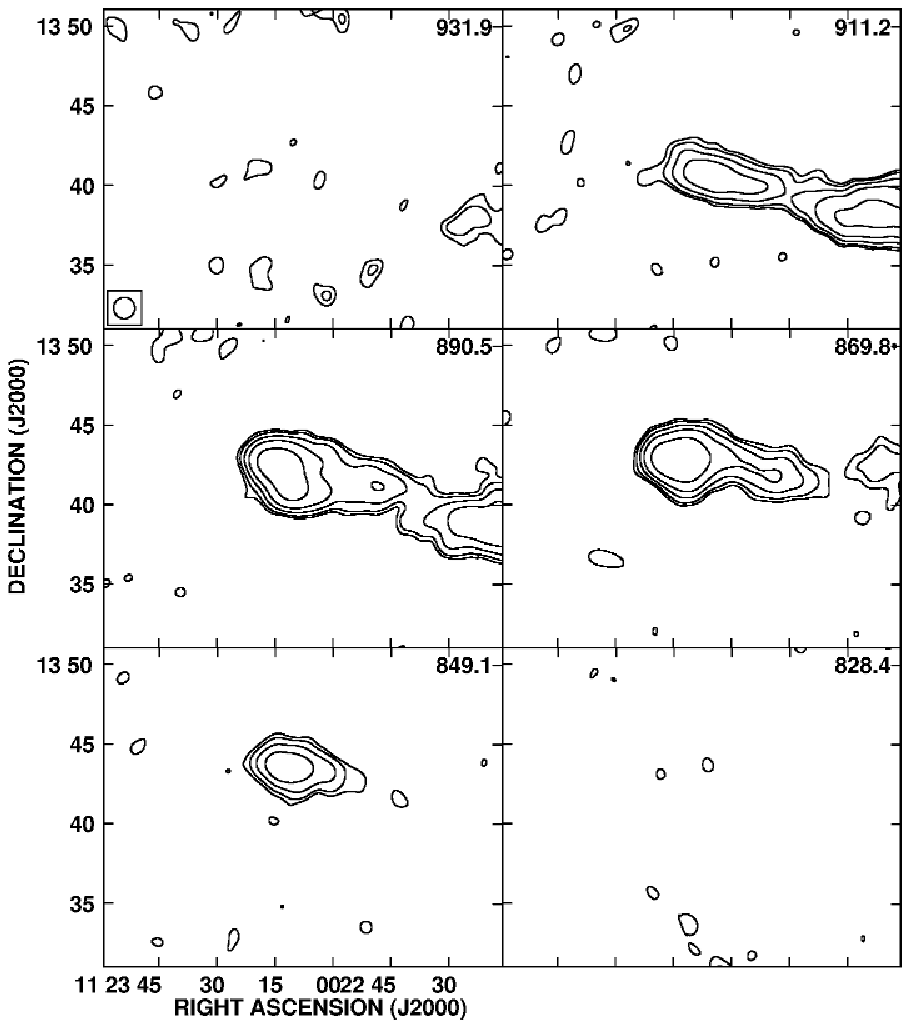}}
\caption{                            
Six-channel map of Leo-TDG. Central velocities (in km\,s$^{-1}$) of each of the channels are written in the upper right corner of each of
the planes. The angular resolution is 80 arcsec; the spectral resolution is 20.7\,km\,s$^{-1}$.
}
\label{3d}
\end{figure}

To illustrate the separation of the two components we made a
contour plot of six channels in which the tail and/or the dwarf system is
visible. This map, included as Fig.~\ref{3d}, shows that the tidal
tail and the dwarf galaxy are separated and there is at least one
channel that only one of them is solely visible.

\section{Discussion}

\subsection{Stellar mass and age}

Due to the very low surface brightness, the SDSS data do not
allow to make a detailed fit to the spectral emission distribution (SED) 
that could be used to 
derive the star formation history and mass of the Leo-TDG.  Still, it is
possible to estimate some information from the photometry. Using
the scaling relation from Bell et al.~\cite{Bell2003}, we can
use {\it g'}\/ and {\it r'}\/ magnitudes and the resulting colour to
get an estimate of the stellar mass.  With {\it g'}\/ = 17.1$^{m}$,
{\it r'}\/ = 16.65$^{m}$, {\it g'}--{\it r'}\/ = 0.45$^{m}$, 
and the values in Table~7 of Bell et
al.~\cite{Bell2003} we calculate the M/L ratio = 3.12. With the
measured L$_{r'}$ of $2.39\times 10^7$ L$_{\odot}$ this results in a
stellar mass of $7.4\times 10^7$ M$_{\odot}$.

A rough limit for the age of the dominant stellar population can
be derived from comparison with the model integrated spectra.
Assuming that the dwarf has at least some more or less recent
star formation (given its large {\rm H}{\sc i} mass), we decided
to use the \textsc{starburst99} code (Leitherer et al.~\cite{leith99,
leith10}; Vazquez et al.~\cite{vazquez}) to model basic
properties of the stellar population of Leo-TDG. 
Independently of the assumed metallicity (2 times solar to 1/20 solar) and 
star formation law (continous or instantanous), for
the measured B$-$V = 0.62 we get a lower age limit of 1 Gyr (which is the
limit of the published models).  If we assume a moderate internal
reddening of 0.3 mag, the age limits are from $3\times10^8$\,yrs for a
star formation burst and solar metallicity to $\approx 10^9$\,yrs
for 20\% of the solar metallicity. Obviously, while being relatively
blue, the majority of the stars formed significantly more
than $10^8$ years ago.  For a more detailed analysis much better
photometry is required.

As estimated by Rots~\cite{rots}, the closest encounter between
NGC\,3627 and NGC\,3628 may have happened $\approx 8 \times
10^{8}$\,yrs ago. Thus most of the stars in Leo-TDG (and
probably the tidal dwarf itself) had to be formed shortly after the
aforementioned collision of these  galaxies. It is not likely
that these stars formed in NGC\,3628 and have been later dragged 
away, as the distance from the parent object is very large.

\subsection{Gas content}
\label{himass}

The gas mass of the Leo-TDG was estimated assuming  
M$_{\mathrm{H}_{\mathrm{I}}}$ [M$_{\odot}$] = 2.36$\times 10^{5}
D^{2}_{\mathrm{Mpc}} \int S_{\nu} \mathrm{d}\nu$, where
$\mathrm{S_{\nu} d\nu}$ is in Jy$\times$\,km\,s$^{-1}$ \citep[van
Gorkom et al.][]{massest}. Using the distance of
12.15\,Mpc  and total flux of 9.0$\pm$0.5\,Jy\,km\,s$^{-1}$ (see
Sect.~\ref{momnt}), we obtained the total mass of the neutral
hydrogen M$_{\mathrm{H}_{\mathrm{I}}} = 3.0$ -- $3.3 \times 10^{8}
\mathrm{M_{\odot}}$. It is somewhat lower than the results from
Stierwalt et al.~\cite{hinew}, but  still of the same 
order of magnitude. The differences are most likely caused by the larger 
beamsize of the Arecibo telescope used by the authors of the former study,
which causes confusion of the emission from the
dwarf candidate by that from the tail.

\subsection{Mass-to-light ratio and the total mass}

The dynamical mass M$_{\mathrm{DYN}}$ of Leo-TDG can be derived
from the rotational velocity at a given radius. For Leo-TDG, the
radial velocity (not corrected for the inclination) gradient is about
35 -- 40\,km\,s$^{-1}$ over some 13\,kpc (with the tail
contribution subtracted).
The neutral hydrogen data do not allow to reliably estimate the
turbulent component. Therefore, we decided to use a conservative
assumption of 10 km\,s$^{-1}$ for a 1-dimensional turbulent
contribution. 
If this is used to calculate the
dynamical mass, one can obtain a total mass of some $7.9\times10^8
\mathrm{M}_{\odot}$. With the inclination unknown, this value can
be treated as a lower limit of the dynamical mass. For a
reasonable inclination of about $60\degr$
(based on the elongation of the the optical and {\rm H}{\sc i}
shape of the dwarf), the total dynamical mass would rise to
M$_{\mathrm{DYN}} = 1.41\times10^9 \mathrm{M}_{\odot}$. 
It should be strongly indicated here, that the dynamical mass estimate 
comes with a large uncertainty. As the dependence of the dynamical mass 
on the (unknown) inclination is given by is given by $M_{DYN} \propto 
1\slash \sin(i)^{2}$, the dynamical mass would largely increase if only Leo-TDG
is a more face-on-orientated system. In general, estimation of masses of the dwarf
galaxies and their distributions is a complicated issue, as even if the 
inclination estimate is proper to some extent, the question of the finite disk
thickness persists \citep[Rhee et al.][]{rhee}. 
The total baryonic content of Leo-TDG can be calculated as a sum
of the stellar mass ($7.4\times 10^7 \mathrm{M}_{\odot}$) and
gaseous component. Assuming a modest estimate of the molecular
gas  mass of $10-30\%$ of the {\rm H}{\sc i} mass \citep[as
M$_{\mathrm{H}_{2}} \slash$M$_{\mathrm{H}_{\mathrm{I}}}$ for
NGC\,3628 is equal to $\approx 20\%$, Obreschkow\&Rawlings][]{obr}, the total gas mass would be
around 3.3 -- 4.3$\times 10^8$ M$_{\odot}$, so the total baryonic
content M$_{\mathrm{BAR}}$ is $4.0$ -- $5.0 \times10^8
\mathrm{M}_{\odot}$.

Estimate of M$_{\mathrm{DYN}}\slash$L$_{\mathrm{B}}$ can also be
derived. L$_{\mathrm{B}}$ [L$_{\odot}$] is equal to
$10^{-0.4\times(\mathrm{M} -\mathrm{M}_{\odot})}$. The B-band
magnitude of the Sun is equal to 5.47 \citep[Cox][]{allen}. This
yields the total B-band luminosity of 2.4$\times
10^7$\,L$_{\odot}$. 
{The M$_{\mathrm{DYN}}\slash$L$_{\mathrm{B}}$
is then 33 -- 59, and
M$_{\mathrm{H}_{\mathrm{I}}}\slash$L$_{\mathrm{B}}$ is 12 -- 14. 

\subsection{Magnetic field}


The resolutions used in our previous study \citep[Nikiel--Wroczy\'nski et
 al.][]{leoeff} -- 4$\arcmin$.3 in the radio continuum and 3$\arcmin$.5 in 
 the {\rm H}{\sc i} data of Stierwalt et al.~\cite{hinew} gave no grounds to
reject the coincidence of the {\rm H}{\sc i} and radio continuum-emitting
regions. There were also no reliable optical images available.   All
this was suggestive for the existence of the magnetic field in Leo-TDG. 

With almost 4 times smaller beam of the {\rm H}{\sc i} data analysed in
this work and using our optical image we could state
that the radio peak is shifted by approximately 1' westwards from the
neutral gas peak and seems to be located outside the optical emission.  A
large fraction of the radio continuum emission may be thus due to a
background source. In the light of the new data we need to revise the
estimate of the magnetic field strength. Setting the upper limit to the
radio emission of 3.0 mJy/beam at the position of gaseous and{
optical feature implies the total magnetic field in the TDG to be B$_{\mathrm{TOT}}
\le$ 2.8\,$\mu$G. The magnetic and cosmic-ray energy density amounts
therefore to P$_{\mathrm{B+CR}}\le 6.8 \times 10^{-13}$ erg$\cdot$cm$^{-3}$.


\subsection{TDG or a non-tidal, LSB galaxy?}

Leo-TDG shares many of its characteristics with the TDGs \citep[Kaviraj et
al.][]{kaviraj}. It is rather bluer than its supposed
progenitor \citep[0.65 compared to 0.8 for NGC\,3628, Paturel et
al.][]{hyperleda}, it is located exactly at the tip of the
tidal tail, and has mass of some $10^{8}\mathrm{M_{\odot}}$,
typical for such  objects.  On the other hand, if identified as
a TDG, the discussed object would be the tidal dwarf most
distant from its parent object, with a calculated
separation of some 140 -- 150\,kpc, while 95\% of the TDG candidates
do not lie more than 20\,kpc from their progenitors \citep[Kaviraj et
al.][]{kaviraj}. Compared to the statistical sample, Leo-TDG
is dim, as it contains less stars than typical TDG candidates.
Among the most distinct features of this galaxy are its low
surface brightness ($\mu_{\mathrm{B}}=25.55$) and very high abundance of
the neutral gas. Because of that, we compare its properties not only with
the TDGs, but also Low Surface  Brightness (LSB) galaxies. We
decided to take a galaxy (F563--1) from the samples collected by de~Blok et
al. \citep{deblok, deblok2} and the "dark" LSB NGC\,3741 ~\citep[Begum
et al.][]{n3741, begum}. As a comparison TDG, we have
chosen the "old TDG" VCC\,2062 \citep[Duc et al.][]{vcc2062}.
The data for the selected objects (TDGs and LSBs) are shown in
Table
\ref{compar}.\\

\begin{table*}[ht!]
\caption{\label{compar} Parameters of Leo-TDG compared to TDG's and LSB's}
\begin{center}
\begin{tabular}{ccccc}
\hline
\hline
Name & Leo-TDG & F563--1 & VCC\,2062 & NGC\,3741\\
Type & TDG & LSB galaxy & TDG & LSB (very dark)\\
Opt. size [kpc] & 7.5 & 3.4$^{1}$ & 0.7$^{1}$ & 1.7\\
{\rm H}{\sc i} size [kpc] & 13 & 16$^{1}$  & 4.2 & 14.6 \\
$\mu_{\mathrm{B}}$ [mag arcsec$^{-1}$] & 25.55 & 23.79 & 24.85 & 24.91\\
B--V & 0.615 & 0.58 & 0.35 & 0.36$^{2}$\\
Total mass [$\mathrm{M_{\odot}}$]& 7.9 -- 14.1$\times 10^{8}$ &$3.9\times 10^{10}$ & 3 -- 4$\times 10^{8}$& $4\times 10^{9}$ \\
Gas content [$\mathrm{M_{\odot}}$] & 3.3 -- 4.3$\times 10^{8}$ & $1.5\times 10^9$  & $0.8\times 10^{8}$ &$1.6\times 10^{8}$ \\
Stellar content [$\mathrm{M_{\odot}}$] & $7.4\times 10^7$ & $2.3\times 10^8$ $^3$ & 0.2 -- 0.7$\times 10^{8}$ & $1.4\times 10^{7}$ \\
M$_{\mathrm{H}_{\mathrm{I}}}\slash$L$_{B}$ & 12 -- 14 & 2.06 & 3 & 6.26 \\
M$_{\mathrm{B}}$ & $-12.97$ & $-16.7$ & $-13$ & $-13.13$\\
M$_{\mathrm{DYN}}\slash$M$_{\mathrm{BAR}}$ & 1.6 -- 3.5 & $\approx$17$^3$ & 2 -- 4 & 24\\
M$_{\mathrm{DYN}}\slash$L$_{\mathrm{B}}$ & 33 -- 59 & 50.1 & $\approx$10 & 149\\
\hline

\hline
\end{tabular}
\end{center}
$^{1}$Derived from its angular size\\
$^{2}$From Taylor and Webster \cite{tw}\\
$^{3}$Derived from the M/L ratio calculated basing on Bell \cite{Bell2003}
\end{table*}

The table clearly shows that the detected galaxy shares
parameters of both TDGs and LSBs.  In fact, it is not the only
one dwarf system that is considered to be either a TDG or an LSB -
likewise is the VCC\,2062 in the Virgo Cluster \citep[Duc et
al.][]{vcc2062}. Both galaxies share similar 
characteristics: they are dim, low--mass systems with low surface
brightness.  They have a significant neutral hydrogen halo,
showing signs of rotation, independent from the tidal arc
movement. The velocity gradients are -- to the limits of
inclination -- similar.  However, the sizes of the {\rm H}{\sc i}
haloes are different, as the one of VCC\,2062 is just 4.2\,kpc
-- approximately three times smaller than that of Leo-TDG. The main
difference between the Leo-TDG and non-tidal LSBs is the dominance
of the gas content in the former one. In most of the LSBs, gas is not a dominant
component: M$_{\mathrm{H}_{\mathrm{I}}}\slash$L$_{\mathrm{B}}$ is
close to 1, and
M$_{\mathrm{H}_{\mathrm{I}}}\slash$M$_{\mathrm{DYN}}$ is less
than 10\% \citep[de Blok][]{deblok}. 
In case of Leo-TDG, {\rm
H}{\sc i} dominates over other fractions, manifesting as a very
high M$_{\mathrm{H}_{\mathrm{I}}}\slash$L$_{\mathrm{B}}$ (12 -- 14) and
M$_{\mathrm{H}_{\mathrm{I}}}\slash$M$_{\mathrm{DYN}}$ of
25 -- 55\%. 
Such high neutral gas content causes
M$_{\mathrm{DYN}}\slash$L$_{\mathrm{B}}$ to be higher than in
VCC\,2062, while M$_{\mathrm{DYN}}\slash$M$_{\mathrm{BAR}}$ is 
very similar. As shown by de
Blok~\citep{deblok2}, non--tidal LSB's have rather high
M$_{\mathrm{DYN}}\slash$M$_{\mathrm{H}_{\mathrm{I}}}$ ratios,
which is different from the case of the gas-dominated Leo-TDG.
M$_{\mathrm{DYN}}\slash$M$_{\mathrm{BAR}}$ of the non-tidal galaxies
are also much higher than 1.6 -- 3.5 estimated for Leo-TDG
\citep[Bournaud et al.][]{bournmass}. All these features strongly
favour scenario of the tidal origin of Leo-TDG.

\section{Summary}

We used the archive VLA {\rm H}{\sc i} spectral data and the SDSS optical observations of the TDG candidate in the Leo Triplet of galaxies.
We obtained maps of the zeroth and first kinematic moments of the {\rm H}{\sc i} content
as well as the distribution of the visible light in the SDSS {\it g', r', i'}\/ bands,
yielding the following results:

\begin{itemize}
 \item There is a massive, star-forming {\rm H}{\sc i} clump at the tip of the tidal tail
 of the Leo Triplet.
 \item Velocity field shows a non-negligible gradient (approximately 35 -- 40\,km\,s$^{-1}$ over 13\,kpc) along the
 N--S (declination) axis, strongly supporting that the detected clump is a self-gravitating,
 tidal dwarf galaxy (TDG).
 \item The dwarf galaxy is unusually distant from its host galaxy (approximately 140 -- 150\,kpc), which is
 more than seven times further than the typical values.
 \item The optical counterpart has been detected in the SDSS {\it g'}\/ and {\it r'}\/ bands. The apparent B magnitude
 $m_{\mathrm{B}}$ is 17.45$^{m}$,  B$-$V is 0.615$^{m}$, and the central surface brightness $\mu_{\mathrm{B}}$ reaches 25.55$^{m}$. 
 The absolute B-band magnitude is $-12.97^{m}$.
 \item The total {\rm H}{\sc i} mass of the clump M$_{\mathrm{H}_{\mathrm{I}}}$ is 3.0 -- 3.3 
 $\times 10^{8} \mathrm{M_{\odot}}$. The stellar content is about $7.4\times 10^7\mathrm{M_{\odot}}$.\\
 \item The stellar population age is $3\times 10^8$ -- $10^9$ years. This means that,
 despite rather blue colour, most of the stars are relatively old.\\
 \item The estimated dynamical mass is just 1.6 -- 3.5 times the baryonic mass. This means that
 the dark matter content plays much less sound role than in the non-tidal dwarf galaxies. Estimated
 M$_{\mathrm{DYN}}\slash$L$_{\mathrm{B}}$ is 33 -- 59, but this value is mostly due to high 
 M$_{\mathrm{H}_{\mathrm{I}}}\slash$L$_{\mathrm{B}}$ (12 -- 14).\\
 \item Compared to the similar objects, Leo-TDG is relatively dim, and has a very high fraction of neutral gas, implying
 a short evolution time and/or low star formation rate since its formation.
 
\end{itemize}

\acknowledgements
We thank the anonymous refferee for valuable comments and suggestions that helped to 
improve our paper.
We acknowledge the usage of the NASA/IPAC Extragalactic Database (NED) which is operated 
by the Jet Propulsion Laboratory, California Institute of Technology, under contract with 
the National Aeronautics and Space Administration. This research has made use of NASA's 
Astrophysics Data System. This research has been supported by the scientific grant a 
the National Science Centre (NCN), decision no. DEC-2011/03/B/ST9/01859.
Funding for the SDSS and SDSS-II has been provided by the Alfred P. Sloan
Foundation, the Participating Institutions, 
the National Science Foundation, the U.S. Department of Energy, the National
Aeronautics and Space Administration, 
the Japanese Monbukagakusho, the Max Planck Society, and the Higher Education
Funding Council for England. 
The SDSS Web Site is http://www.sdss.org/.


\begin{thebibliography}

\bibitem[2009]{DR7} Abazajian, K. N.; Adelman-McCarthy, J. K.; Ag\"ueros, M. A. et al. 2009, \apjs, 182, 543

\bibitem[2003]{Bell2003} Bell, E.~F; McIntosh, D.~H; Katz, N. et al. 2003, \apjs, 149, 289

\bibitem[2005]{n3741} Begum, A.; Chengalur, J.~N.; Karachentsev, I.~D.2005, \aap, 433, 1

\bibitem[2008]{begum} Begum, A.; Chengalur, J.~N.; Karachentsev, I.~D. 2008, \mnras, 386, 1667

\bibitem[2004]{brinks} Brinks, E.; Duc, P.-A.; Walter, F.2004, in: Proceedings of the International Astronomical Union, 2004, 532

\bibitem[2010]{bournmass} Bournaud, F. 2010, Advances in Astronomy, 2010

\bibitem[2008]{bournrev} Bournaud, Duc 2008, ApJ, 672, 787

\bibitem[1998]{NVSS} Condon, J.~J.; Cotton, W.~D.; Greisen, E.~W. et al. 1998, \apjl, 115, 1693

\bibitem[1998]{allen} Allen's astrophysical quantities, 4th ed. Publisher: New York: AIP Press;
Springer, 2000. Edited by Arthur N. Cox. ISBN: 0387987460

\bibitem[1995]{deblok} de Blok, W. J. G.; van der Hulst, J. M.; Bothun, G. D. 1995, \mnras, 274, 235

\bibitem[1996]{deblok2} de Blok, W. J. G.; McGaugh, S. S.; van der Hulst, J. M. 1996, \mnras, 283, 218

\bibitem[2007]{vcc2062} Duc, P.-A.; Braine, J.; Lisenfeld, U. et al. 2007, \aap, 475, 187

\bibitem[2008]{bournvirg} Duc, P.-A; Bournaud, F.; Brinks, E. 2008, IAU Symposium, 244, 216

\bibitem[1979]{arecibo} Haynes, M.~P.; Giovanelli, R.; Roberts, M.~S. 1979, \apj, 229, 83 

\bibitem[2000]{Hunter2000} Hunter, D.~A.; Hunsberger, S.~D.; Roye, E.~W. 2000, \apj, 542, 137

\bibitem[2005]{jester} Jester, S.; Schneider, D.~P.; Richards, G.~T. 2005, \aj, 130, 873

\bibitem[2012]{kaviraj} Kaviraj, S.; Darg, D.; Lintott, C. et al. 2012, \mnras, 419, 70

\bibitem[1974]{korm} Kormendy, J.; Bahcall, J.~N. 1974, \aj, 79, 671

\bibitem[1999]{leith99} Leitherer, C.; Schaerer, D.; Goldader, J.~D. 1999, \apjs, 123, 3

\bibitem[2010]{leith10} Leitherer, C; Ortiz Ot\'alvaro, P.~A.; Bresolin, F. 2010, \apjs, 189, 309

\bibitem[2002]{Makarova2002} Makarova, L.~N.; Grebel, E.~K.; Karachentsev, I.~D. et al. 2002, \aap, 396, 473

\bibitem[2005]{vhi21} Minchin, R.; Davies, J.; Disney, M.et al. 2005, \apjl, 622, 21

\bibitem[1992]{mirabel} Mirabel, I.~F.; Dottori, H.; Lutz, D. 1992, \aap, 256, 19

\bibitem[2011]{miskolczi} Miskolczi, A.; Bomans, D. J.; Dettmar, R.-J. 2011, \aap, 536, A66

\bibitem[2013]{leoeff} Nikiel-Wroczy\'nski, B.; Soida, M.; Urbanik, M. et al. 2013, \aap, 553, A4

\bibitem[2009]{obr} Obreschkow, D.; Rawlings, S. 2009, \mnras, 394, 1857

\bibitem[2003]{hyperleda} Paturel, G.; Petit, C.; Prugniel, Ph. et al. 2003, \aap, 412, 45

\bibitem[2004]{rhee} Rhee, G.; Valenzuela, O.; Klypin, A. et al. 2004, \apj, 617, 1059

\bibitem[1978]{rots} Rots, A.~H. 1978, \aj, 83, 219

\bibitem[1978]{schweizer} Schweizer, F. 1978. In Structure and Properties of Nearby Galaxies, 
ed. E.M. Berkhuijsen, R. Wielebinski, p. 279 

\bibitem[2001]{3627} Soida, M.; Urbanik, M.; Beck, R. et al. 2001, \aap, 378, 40

\bibitem[2009]{hinew} Stierwalt, S.; Haynes, M.~P.; Giovanelli, R. et al. 2009, \apj, 138, 338

\bibitem[2005]{tw} Taylor, E.~N.; Webster, R.~L. 2005, \apj, 634, 1067

\bibitem[1986]{massest} van Gorkom, J.~H.; Knapp, G.~R.; Raimond, E. et al. 1986, \aj, 91, 791

\bibitem[2005]{vazquez} V\'azquez, G.~A.; Leitherer, C. 2005, \apj, 621, 695

\bibitem[2012]{3627dwarf} We\.zgowiec, M.; Soida, M.; Bomans, D.~J. 2012, \aap, 544, 113

\bibitem[1956]{zwicky} Zwicky, F.; 1956, Ergebnisse der Exakten Naturwissenschaften 29, 344

\end{thebibliography}
\end{document}